\documentstyle[11pt,newpasp,twoside]{article}
\def\edcomment#1{\iffalse\marginpar{\raggedright\sl#1\/}\else\relax\fi}
\reversemarginpar

\begin{document}
\title{Pulsating White Dwarfs in Cataclysmic Variables: \\ The Marriage of ZZ Cet and Dwarf Nova}
 \author{Brian Warner and Patrick A. Woudt}
\affil{Department of Astronomy, University of Cape Town, Rondebosch 7700, South Africa}

\begin{abstract}
There are now four dwarf novae known with white dwarf primaries that show large amplitude
non-radial oscillations of the kind seen in ZZ Cet stars. We compare the properties of these
stars and point out that by the end of the Sloan Digital Sky Survey more than 30 should be known.
\end{abstract}

\section{Introduction}

The DA white dwarf non-radially pulsators stars, known as ZZ Cet stars, 
have traditionally been found as isolated field objects. Until recently 34 were 
known (Fontaine et al.~2003), but a further tranche of 34 were just 
announced (D.~Sullivan, this conference). The ZZ Cet pulsation strip in the HR diagram 
roughly extends over the range 11\,000 K $<$ $T_{eff}$ $<$ 12\,000 K (Bergeron et 
al.~1995), so it is of interest to see whether cataclysmic variables (CVs) with 
temperatures in that range also show pulsations (though the width and centre 
of gravity of the instability strip may be modified in the presence of 
accretion: Townsley \& Bildsten 2003).

   Until recently only one such hybrid CV/ZZ system was known: GW Lib 
(Warner \& van Zyl 1998), which has a measured $T_{eff}$ of 14\,700 K (Szkody et 
al.~2002a) -- though this may be affected by non-allowance for the mysterious 
emission source that adds a 2.08 h photometric modulation independent of 
the 1.28 h orbital period (Woudt \& Warner 2002). The principal periods in 
GW Lib are near 650 s, 376 s and 236 s. 

     A second CV/ZZ system, SDSS\,J161033.64-010223.3 (SDSS\,1610 hereafter), was discovered in June 
2003 (Woudt \& Warner 2003) and was selected as a candidate on the basis 
of its spectrum as published in the first release of CVs in the Sloan Digital 
Sky Survey (Szkody et al.~2002b). SDSS\,1610 resembles GW Lib in clearly 
showing absorption lines of the underlying white dwarf primary, as well as 
the emission lines characteristic of an accreting system. Its light curve is 
shown in Figure 1 (top panel). Its periodicities are near 607 s, 345 s and 221 s, with a 
harmonic at 304 s.

   There is evidently a window of opportunity among the CVs of low rate of 
accretion ($\dot{M}$): the $T_{eff}$ of the white dwarf is determined by $\dot{M}$, largely 
through compressional heating in the interior (Sion 2003), and it happens 
that an $\dot{M}$ that maintains $T_{eff}$ in the instability strip is sufficient to produce 
Balmer emission lines but not too large to give the accretion disc a 
luminosity large enough to hide the flux from the white dwarf. From the 
observed depth of, e.g., H$\beta$, the fraction of flux contributed by the white 
dwarf can be estimated; for GW Lib this is $\sim$ 50\% and in SDSS 1610 it is 
even greater.

   The low $\dot{M}$ in these systems automatically leads to the expectation that 
they will be dwarf novae of very long outburst interval $T_{out}$ (see, e.g., 
Warner (1995)). Comparison with Z Cha, which has $T_{out} \sim$ 50 d, shows that 
although the latter's primary is visible in the spectrum, it is far more covered 
by accretion disc flux than in the known CV/ZZ stars. At the other end of the 
$T_{out}$ range, GW Lib has only had one observed outburst (in 1983) and SDSS 
1610 has had none. Because the low $\dot{M}$ systems are intrinsically faint 
they will in general be apparently faint -- it is the ability of the SDSS to find 
CVs to faint limits that has opened the possibility of increasing the number 
of known CV/ZZ stars.

\section{Pulsation frequencies}

    Some of the ZZ Cet stars show frequency patterns that have been ascribed 
to direct resonance between the principal driving modes and the wealth of 
other available modes (O'Donoghue, Warner \& Cropper 1992). Thus in VY 
Hor, GD 154 and PG 1351+489 the eigenfrequencies are almost completely 
described by the sequence $nf$ and $(m + {1\over{2}} + \epsilon)\,f$, where $n$ = 1, 2, ... and $m$ = 0, 
1, 2, ..., and $\epsilon$ is a small quantity (O'Donoghue, Warner \& Cropper 1992; 
Robinson et al.~1978; Winget, Nather \& Hill 1987). For VY Hor $\epsilon$ = 0.037, 
for GD 154 $\epsilon$~=~--0.03 and for PG 1351 $\epsilon$ = --0.03. In GW Lib and SDSS\,1610, 
with the still rather limited observational data for the latter, we find a similar 
situation, but with the sequences $nf$ and $(m + {3\over{4}} + \epsilon)\,f$, where $\epsilon \approx -0.03$ for 
GW Lib and $\epsilon \approx 0.00$ for SDSS\,1610. This may be a coincidence, but it 
could indicate that a different resonance condition is operating in accreting 
ZZ Cet stars.

\begin{figure}[t]
\plotfiddle{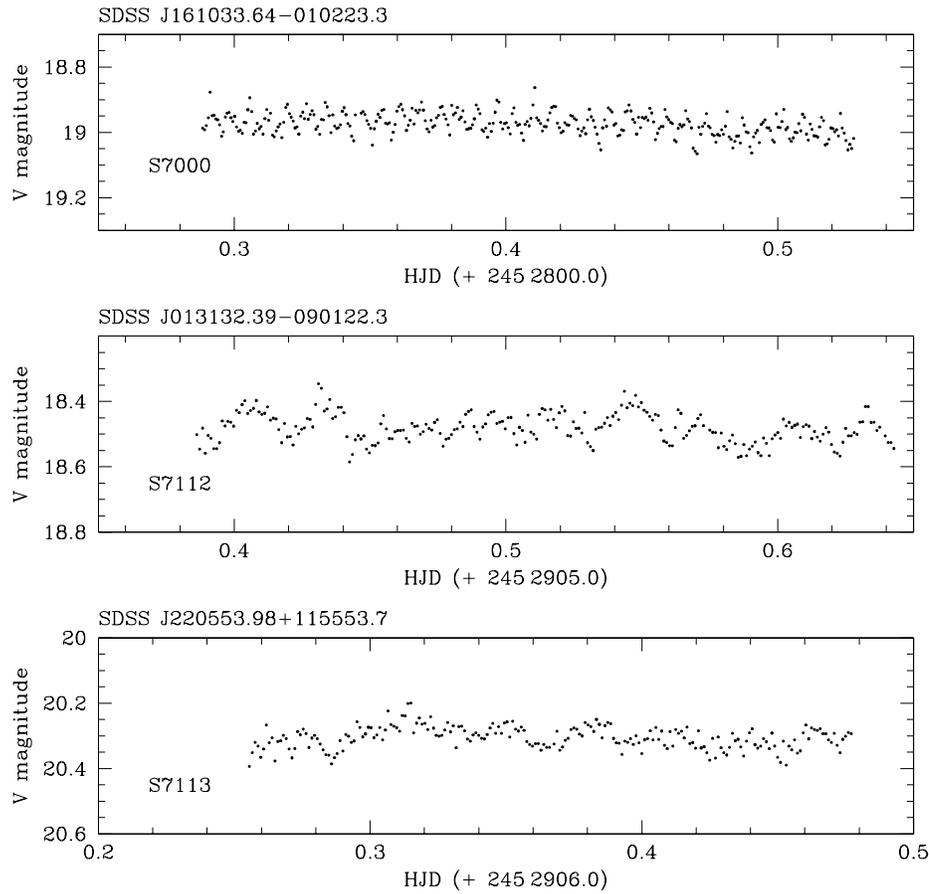}{11.0cm}{0}{65}{65}{-200}{-119}
\caption{Light curves of the three new CV/ZZ stars, obtained with the University of Cape Town
CCD and the 1.0-m and 1.9-m telescopes at the South African Astronomical Observatory.}
\label{warnerfig1}
\end{figure}

\section{Further discoveries}

    The publication of the second SDSS CV release (Szkody et al.~2003) 
revealed at least three further candidate CV/ZZ stars. We have observed 
SDSS\,J013132.39-090122.3 and SDSS\,J220553.98+115553.7 (see also Figure 1) and find that 
indeed both have ZZ Cet primaries. At the time of writing this leaves the 
strong candidate SDSS 1238 to be examined later in the observing season.

    The Fourier transforms in Figure 2 show that in SDSS\,0131 the dominant periods are near 595 s and 335 s with weak 
oscillations also near 260 s; SDSS\,2205 has strong oscillations near 575 s 
and 330 s, with weak oscillations near 475 s. In both cases the ratio of the 
stronger periods is $\sim$ 1.75, as in GW Lib and SDSS\,1610.

    The total number of CVs in the first two SSDS releases is 60, of which we 
have found that probably $\sim$ 8\% are CV/ZZ combinations. It is estimated that 
the final total of CVs found in the SDSS will be $\sim$ 400 (Szkody et al.~2002b), 
so in a few years we may expect to have more than 30 CV/ZZ stars to study 
in detail.

\begin{figure}[t]
\plotfiddle{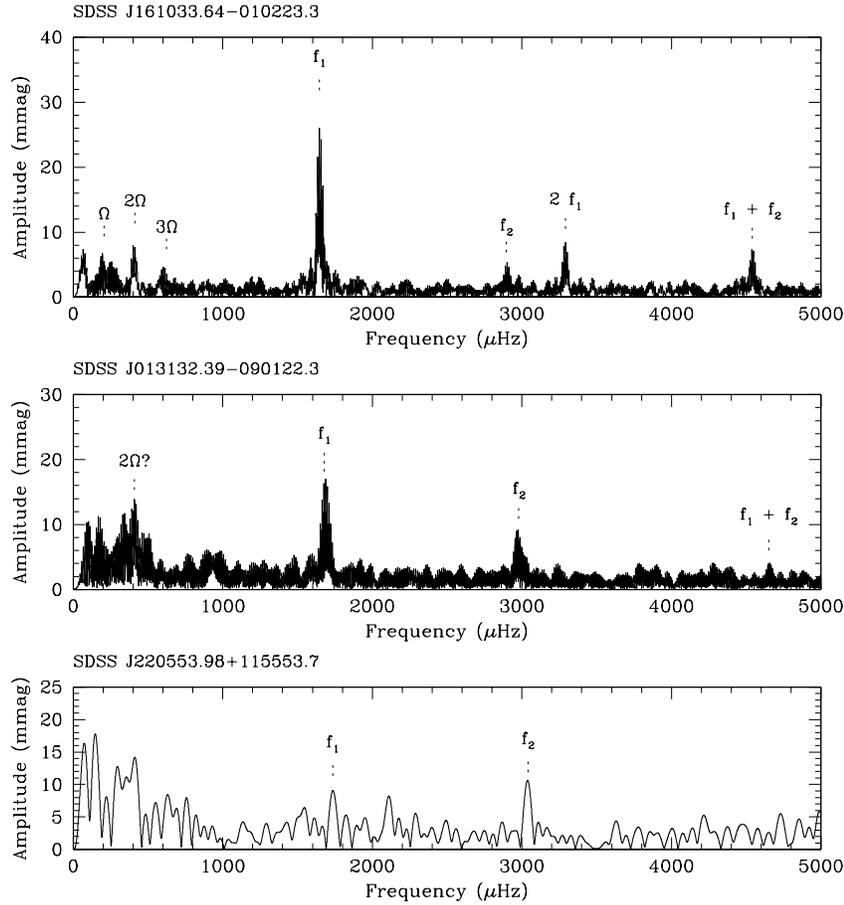}{11.3cm}{0}{70}{70}{-210}{-145}
\caption{Fourier transforms of the three new CV/ZZ stars. The FTs of SDSS\,1610 and SDSS\,0131 are based on three (SDSS\,1610) 
and four (SDSS\,0131) nights of data, respectively; the FT of SDSS\,2205 is based on a single 5.5-h observing run (run S7113 
shown in Figure 1), but the same frequencies are seen 
during other runs on other nights. The most prominent frequencies are labelled with $f_1$ and $f_2$.}
\label{warnerfig2}
\end{figure}

\bigskip
\acknowledgements{BW is supported by funds from the University of Cape Town;
PAW is supported by strategic funds made available to BW by the University
of Cape Town and by the National Research Foundation.}


\begin{references}
\reference Bergeron, P., Wesemael, F., Lamontagne, R., Fontaine, G., 
  Saffer, R.A., \& Allard, N.F. 1995, \apj, 449, 258
\reference Fontaine, G., Bergeron, P, Bill\`eres, M., \& Charpinet, S. 2003,
  \apj, 591, 1184
\reference O'Donoghue, D., Warner, B., \& Cropper, M. 1992, \mnras, 258, 415
\reference Robinson, E.L., Stover, R.J., Nather, R.E., \& McGraw, J.T. 1978,
  \apj, 220, 614
\reference Sion, E.M. 2003, Proc.~13th European Workshop on White Dwarfs (Naples)
\reference Szkody, P., Gaensicke, B.T., Howell, S.B., \& Sion, E.M. 2002a,
  \apj, 575, 79
\reference Szkody, P., et al. 2002b, \aj, 123, 430
\reference Szkody, P., et al. 2003, \aj, 126, 1499
\reference Townsley, D.M., \& Bildsten, L. 2003, \apj, in press
\reference Warner, B. 1995, Cataclysmic Variable Stars, Cambridge University
  Press
\reference Warner, B., \& van Zyl, L. 1998, IAU Symp. No. 185, 321
\reference Winget, D.E., Nather, R.E., \& Hill, J.A. 1987, \apj, 316, 305
\reference Woudt, P.A., \& Warner, B. 2002, \apss, 282, 433
\reference Woudt, P.A., \& Warner, B. 2003, \mnras, in press
\end{references}
\end{document}